\documentclass[prd,twocolumn]{revtex4}

\usepackage[latin1]{inputenc}
\usepackage{graphicx}
\usepackage{epsfig}
\usepackage{bm}
\usepackage{amssymb}
\usepackage{float}
\usepackage{amsmath}
\usepackage{dcolumn}
\usepackage{cancel}
\usepackage{color}
\usepackage{epstopdf}
\usepackage{nccbbb}
\usepackage{lipsum}
\usepackage{orcidlink}

\begin{document}

\title{Observational constraints on phenomenological emergent dark energy and barotropic dark matter characterized by a constant equation of state parameter}

\author{Jian-Qi Liu\orcidlink{0009-0005-7615-0149}}
\author{Yan-Hong Yao\orcidlink{0000-0001-5283-3635}}
\email{yaoyh29@mail.sysu.edu.cn}
\author{Yan Su\orcidlink{0009-0003-1993-0769}}
\author{Jun-Chao Wang\orcidlink{0000-0002-4099-9520}}
\author{Jia-Wei Wu}

\affiliation{School of Physics and Astronomy, Sun Yat-sen University, 2 Daxue Road, Tangjia, Zhuhai, 519082, People's Republic of China}

\begin{abstract}
    While cold dark matter is well-supported by a broad array of cosmological data alongside a cosmological constant, it faces several inherent challenges at small scales. These challenges have spurred the exploration of various alternative dark matter candidates beyond cold dark matter, leaving the question of "What is DM?" relatively open. Therefore, we propose a new cosmological model that considers dark matter as a barotropic fluid with a constant equation of state parameter and interprets dark energy as the phenomenological emergent dark energy rather than a cosmological constant. This proposal is based on extensive research on the extended properties of dark matter in the context of a cosmological constant and the intriguing findings that have emerged from our exploration of dark matter properties within the context of PEDE in our previous studies. We then place constraints on this model in light of the Planck 2018 Cosmic Microwave Background (CMB) anisotropies, baryon acoustic oscillation (BAO) measurements, the Pantheon compilation of Type Ia supernovae, a prior on $H_0$  that based on the latest local measurement by Riess et al., and the combination of KiDS and the VISTA Kilo-Degree Infrared Galaxy Survey (KiDS+VIKING-450). The results indicate a preference for a positive dark matter equation of state parameter at 68\% confidence level for CMB+BAO, CMB+BAO+Pantheon and CMB+BAO+Pantheon+$H_0$ datasets. Furthermore, the Hubble tension between all of the datasets we used with R22 is very close to those of the PEDE, and the $S_8$ tension between Planck 2018 and KiDS+VIKING-450 is reduced from 2.3$\sigma$ in the PEDE model to 0.4$\sigma$ in the new model. However, Bayesian evidence indicates that PEDE favors our new model with very strong evidence from all the datasets considered in this study. Consequently, we conclude that the PEDE+$w_{\rm dm}$ model is not a viable alternative to the PEDE model.
\end{abstract}

\maketitle

\section{introduction}\label{intro}
Dark matter (DM), a mysterious component of the universe that does not interact with photons, is expected to account for one-fourth of the energy budget of the universe today. Although the nature of DM remains uncertain, we currently have a simple and popular DM candidate, i.e., the cold dark matter (CDM). In the standard $\Lambda$-Cold Dark Matter ($\Lambda$CDM) model, CDM is considered as a non-interacting perfect fluid having zero equation of state (EoS) parameter as well as zero sound speed and zero viscosity. Accompanied by a cosmological constant, CDM is supported by a wide range of cosmological data~\citep{riess1998observational,perlmutter1999measurements,dunkley2011atacama,hinshaw2013nine,story2015measurement,alam2017clustering,troxel2018dark,aghanim2020planck-1}. However, CDM loses its magic at small scales, facing issues inherent in this paradigm, such as the missing satellite~\citep{klypin1999missing, moore1999dark}, too-big-to-fail~\citep{boylan2012milky}, and core-cusp problem~\citep{moore1999cold, springel2008aquarius}. Small-scale issues of the CDM paradigm have motivated many new DM candidates beyond CDM, including warm DM~\cite{blumenthal1982galaxy,bode2001halo}, fuzzy DM~\cite{hu2000fuzzy,marsh2014model}, interacting DM~\cite{spergel2000observational} and decaying DM~\cite{wang2014cosmological}. These new candidates are proposed to suppress the formation of low-mass structures and to be consistent with large-scale data simultaneously.

Inspired by the limitations of CDM and the diversity of DM candidates, in this article, we will investigate the extended properties of DM through the generalized dark matter (GDM) framework. The GDM first proposed in~\cite{hu1998structure} (for follow-up studies on the GDM model, we recommend taking a look at ~\cite{mueller2005cosmological, kumar2014observational, armendariz2014cold, kopp2018dark, kumar2019testing, ilic2021dark, pan2023iwdm}) and is specified by the DM EoS, the rest-frame sound speed, and the viscosity. Since there have already been many researches on the extended properties of DM in the context of a cosmological constant~\cite{mueller2005cosmological,kopp2018dark,kumar2019testing,ilic2021dark,armendariz2014cold}, in this work we will follow the approach of Ref.~\citep{yao2024observational}, i.e., considering the phenomenological emergent dark energy (PEDE) as the invisible fuel that accelerates the expansion rate of the current universe. The PEDE model, initially proposed by the authors of Ref.\cite{li2019simple} to address the Hubble tension, has been subsequently extended to a generalized parameterization form capable of accommodating both the cosmological constant and the PEDE model\cite{li2020evidence} (also see~\cite{yang2021generalized,hernandez2020generalized}). Notably, the PEDE model features an identical number of free parameters as the spatially flat $\Lambda$CDM model. This framework was further examined in Ref.~\cite{pan2020reconciling}, which explored its complete evolution encompassing both background and perturbations, yielding results consistent with previous works regarding the $H_0$ tension by using recent observational datasets. In the Ref.~\citep{yao2024observational}, the authors set the DM EoS parameter as a free parameter and the DM sound speed to zero in the context of PEDE. After placing constraints on this model with some specific combinations of the Planck 2018 Cosmic Microwave Background (CMB) anisotropies measurements, baryon acoustic oscillation (BAO) measurements, and the Pantheon compilation of Type Ia supernovae, the final results indicate a preference for a negative DM EoS parameter at $95\%$ CL for CMB+Pantheon and CMB+BAO+Pantheon datasets, which is very interesting and worth further investigation. In order to investigate whether introducing other GDM parameters in the context of PEDE would lead to different interesting results, in this paper, building upon the assumption of keeping the DM EoS parameter as a free parameter, we further assume that the DM sound speed is a parameter and hence not being fixed at zero. For the sake of model simplicity, we assume that the DM non-adiabatic sound speed and viscosity are zero, i.e., DM is barotropic. Then we will constrain the new model using CMB, BAO, Ia supernovae (SN Ia), and weak lensing (WL) observational data, and analyze the fitting results of relevant free parameters to see if there are indeed some interesting outcomes emerging.

This paper is organized as follows. Section~\ref{methods} outlines the key equations of the PEDE+$w_{\mathrm{dm}}$ model. In section~\ref{data}, we present the observational datasets and the statistical methodology. In section~\ref{results}, we present the observational constraints and implications of the PEDE + $w_{\mathrm{dm}}$ model. In the last section, we make a brief conclusion to this paper.

\section{Review of the PEDE+$w_{\mathrm{dm}}$ Model}\label{methods}
In this work, we focus on a spatially flat, homogeneous, and isotropic spacetime described by the spatially flat Friedmann-Robertson-Walker (FRW) metric. Additionally, we assume that the gravitational sector of the universe is suitably described by general relativity, where matter is minimally coupled to gravity. Furthermore, we assume that none of the fluids interact non-gravitationally with each other, and the universe consists of radiation, baryons, GDM, and PEDE. Consequently, we can express the Hubble parameter as
\begin{align}
    \frac{H^{2}}{H_{0}^{2}} = & \hspace{0.1cm} \Omega_{\mathrm{r0}}(1+z)^{4} + \Omega_{\mathrm{dm0}}(1+z)^{3(1+w_{\mathrm{dm}})} \notag \\
                              & + \Omega_{\mathrm{b0}}(1+z)^{3} + \Omega_{\mathrm{de}}(z),
    \label{Eq:H}
\end{align}
where $H$ is the Hubble parameter, $\Omega_{\mathrm{r0}}$, $\Omega_{\mathrm{dm0}}$, $\Omega_{\mathrm{b0}}$, and $\Omega_{\mathrm{de}}$ are the density parameters for radiation, GDM, baryons, and PEDE respectively, here $\Omega_{\mathrm{de}}$ is parameterized in the following form~\cite{li2019simple,pan2020reconciling}:
\begin{align}
    \Omega_{\mathrm{de}}(z) = \Omega_{\mathrm{de0}}[1+\tanh(\log_{10}(1+z))].
    \label{Eq:Omde}
\end{align}
where $\Omega_{\rm de0}=1-\Omega_{\rm r0}-\Omega_{\rm dm0}-\Omega_{\rm b0}$ and $1+z=a^{-1}$ (note that we have set the current
value of the scale factor to be unity). Since we assume that none of the fluids interact non-gravitationally with each other, the PEDE conservation equation reads
\begin{align}
    \dot{\rho_{\mathrm{de}}}(z) + 3H(1+w_{\mathrm{de}}(z))\rho_{\mathrm{de}}(z)=0.
    \label{Eq:DEcon}
\end{align}
Here, an over dot represents the derivative with respect to cosmic time. From the equation above, one can derive a following relation between PEDE's EoS and density:
\begin{align}
    w_{\mathrm{de}}(z) = -1 + \frac{1}{1+z} \times \frac{\mathrm{d} \ln{\Omega_{\mathrm{de}}(z)}}{\mathrm{d}z},
\end{align}
substituting Eq.~(\ref{Eq:Omde}) in it, we obtain an explicitly PEDE EoS as follows~\cite{li2019simple,pan2020reconciling}
\begin{align}
    w_{\mathrm{de}}(z) = -1 - \frac{1}{3\ln10} \times [1+\tanh(\log_{10}(1+z))].
\end{align}
It can be inferred from the above equation that the PEDE EoS exhibits an intriguing symmetrical feature. Specifically, for the distant past, i.e., $z \rightarrow \infty$, one obtains $w_{\mathrm{de}} \rightarrow -1-\frac{2}{3\ln10}$. For the distant future, i.e., $z \rightarrow -1$, one finds $w_{\mathrm{de}} \rightarrow -1$. At the present time when $z = 0$, we observe that $w_{\mathrm{de}} = -1-\frac{1}{3\ln10}$, indicating a phantom dark energy (DE) EoS. As briefly described in Ref.~\citep{li2019simple}, the pivot point of transition for the PEDE EoS can be considered to be the redshift at which matter-dark energy densities are equal.

In the conformal Newtonian gauge, the perturbed FRW metric is expressed in the following form:
\begin{align}
    ds^2=a^2(\tau)[-(1+2\psi)d\tau^2+(1-2\phi)d\vec{r}^2],
\end{align}
where $\psi$ and $\phi$ represent the metric potentials, while $\vec{r}$ denotes the three spatial coordinates. By considering the first order perturbed part of the conserved stress-energy momentum tensor, one can derive the following continuity and Euler equations (in Fourier space) for GDM and PEDE~\citep{kumar2019testing}:
\begin{eqnarray}
    \label{Eq:pertur-1}
    \delta_{\rm ds}^{\prime}&=& - (1+w_{\rm ds}) \left(\theta_{\rm ds}-3\phi^{\prime}\right)- 3 \mathcal{H} \left(\frac{\delta p_{\rm ds}}{\delta\rho_{\rm ds}} - w_{\rm ds}
    \right)\delta_{\rm ds}\hspace{0.5cm}\\
    \theta_{\rm ds}^{\prime} &=&-\mathcal{H}
    (1-3c_\mathrm{\rm ad,ds}^2)\theta_{\rm ds}  + \frac{\delta p_{\rm ds}/\delta\rho_{\rm ds}}{1+w_{\rm ds}}k^2\delta_{\rm ds} + k^2\psi
\end{eqnarray}
Here, a prime stands for the conformal time derivative, $\mathcal{H}$ is the conformal Hubble parameter, and k is the magnitude of the wavevector $\vec{k}$. $\delta_{\rm ds}$ and $\theta_{\rm ds}$ are the relative density and velocity divergence perturbations of the dark sector (DS:DM or DE), $w_{\rm ds}$ and $c_{\rm ad,ds}^2=\frac{p_{\rm ds}^{\prime}}{\rho_{ds}^{\prime}}=w_{\rm ds}-\frac{w_{\rm ds}^{\prime}}{3\mathcal{H}\left(1+w_{\rm ds}\right)}$ denote the DS EoS and the square of DS adiabatic sound speed, respectively. And $\frac{\delta p_{\rm ds}}{\delta\rho_{\rm ds}}$ is the DS sound speed in the Newtonian gauge, it can be expressed as:
\begin{equation}
    \label{deltaP}
    \frac{\delta p_{\rm ds}}{\delta\rho_{\rm ds}}=c_\mathrm{\rm s,ds}^2+3\mathcal{H}\left(1+w_{\rm ds}\right)\left(c_\mathrm{\rm s,ds}^2-c_\mathrm{\rm ad,ds}^2\right)\frac{\theta_{\rm ds}}{k^2},
\end{equation}
here $c_\mathrm{\rm s,ds}^2=c_\mathrm{\rm ad,ds}^2+c_\mathrm{\rm nad,ds}^2$ is the square of DS sound speed in the rest frame. And $c_\mathrm{\rm nad,ds}^2$ is the square of DS non-adiabatic sound speed, it describes the micro-scale properties of DS and needs to be provided independently. In this work, we consider $c_\mathrm{\rm nad,dm}^2=0$
(Therefore $c_\mathrm{\rm s,dm}^2=c_\mathrm{\rm ad,dm}^2=w_{\rm dm}$) and $c_\mathrm{\rm s,de}^2=1$, then the continuity and Euler equations for GDM and PEDE can be rewritten as:
\begin{eqnarray}
    \label{Eq:pertur-2}
    \delta_{\rm dm}^{\prime}&=& - (1+w_{\rm dm}) \left(\theta_{\rm dm}-3\phi^{\prime}\right)\\
    \theta_{\rm dm}^{\prime} &=&-\mathcal{H}(1-3w_{\rm dm})\theta_{\rm dm}  + \frac{w_{\rm dm}}{1+w_{\rm dm}}k^2\delta_{\rm dm} + k^2\psi\hspace{0.5cm}
\end{eqnarray}
\begin{align}
    \dot{\delta}_{\mathrm{de}}= & -(1+w_{\mathrm{de}}) \left(\theta_{\mathrm{de}} - 3 \dot{\phi} \right) -3 \mathcal{H}(1 - w_{\mathrm{de}})\delta_{\mathrm{de}} \notag                           \\
                                & -3\mathcal{H}w_{\mathrm{de}}^{\prime}\frac{\theta_{\mathrm{de}}}{k^2}- 9 (1+w_{\mathrm{de}})(1- w_{\mathrm{de}})\mathcal{H}^2 \frac{\theta_{\mathrm{de}}}{k^2}, \\
    \dot{\theta}_{\mathrm{de}}= & 2\mathcal{H} \theta_{\mathrm{de}}  + \frac{1}{1+w_{\mathrm{de}}}k^2 \delta_{\mathrm{de}} + k^2\psi.
\end{align}
Having presented the equations above, the background and perturbation dynamics of our new model (label as PEDE+$w_{\rm dm}$) is clearly understood.

\begin{figure*}
    \centering
    \begin{minipage}{0.45\linewidth}
        \centerline{\includegraphics[width=1\textwidth]{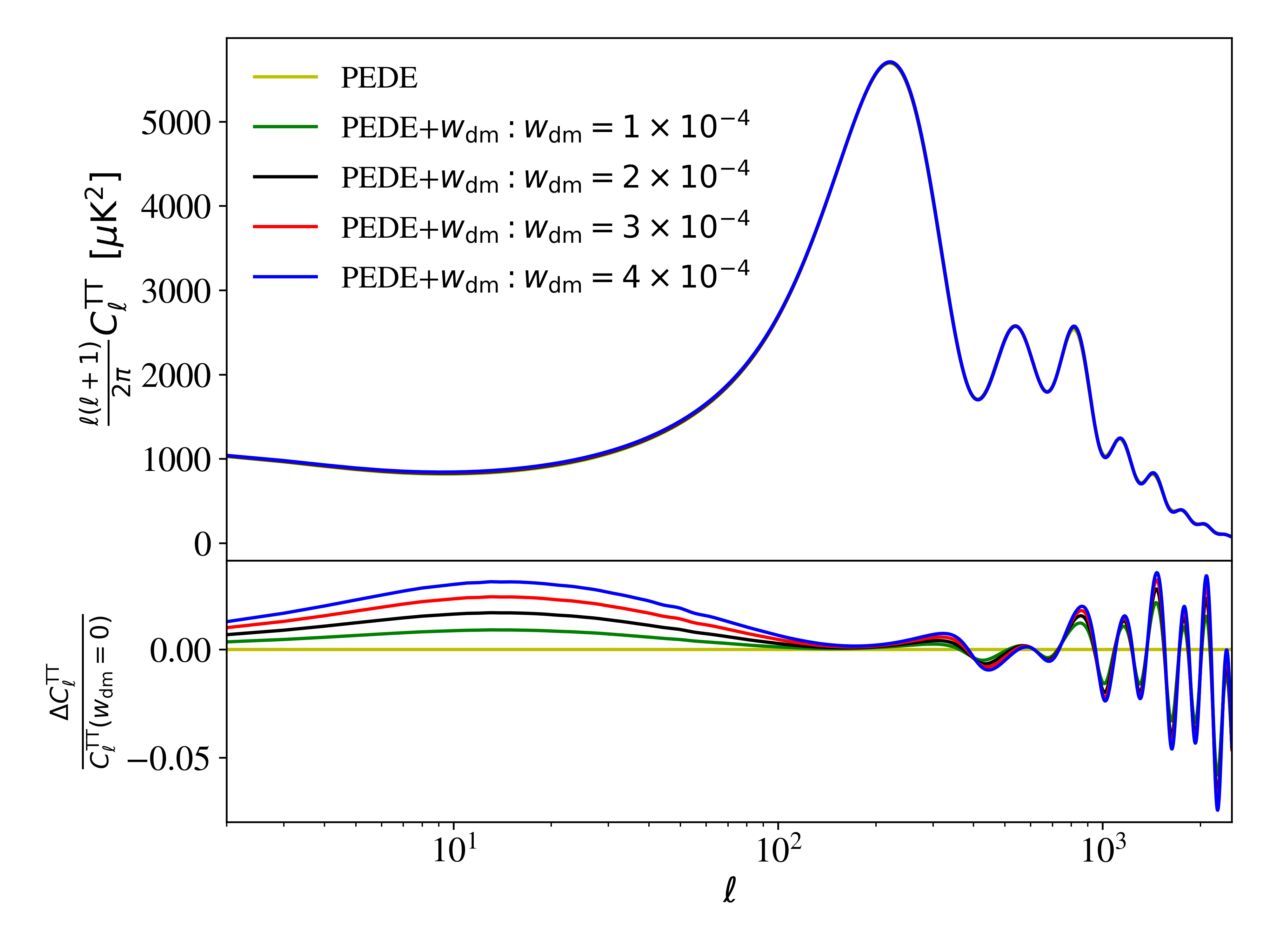}}
    \end{minipage}
    \begin{minipage}{0.45\linewidth}
        \centerline{\includegraphics[width=1\textwidth]{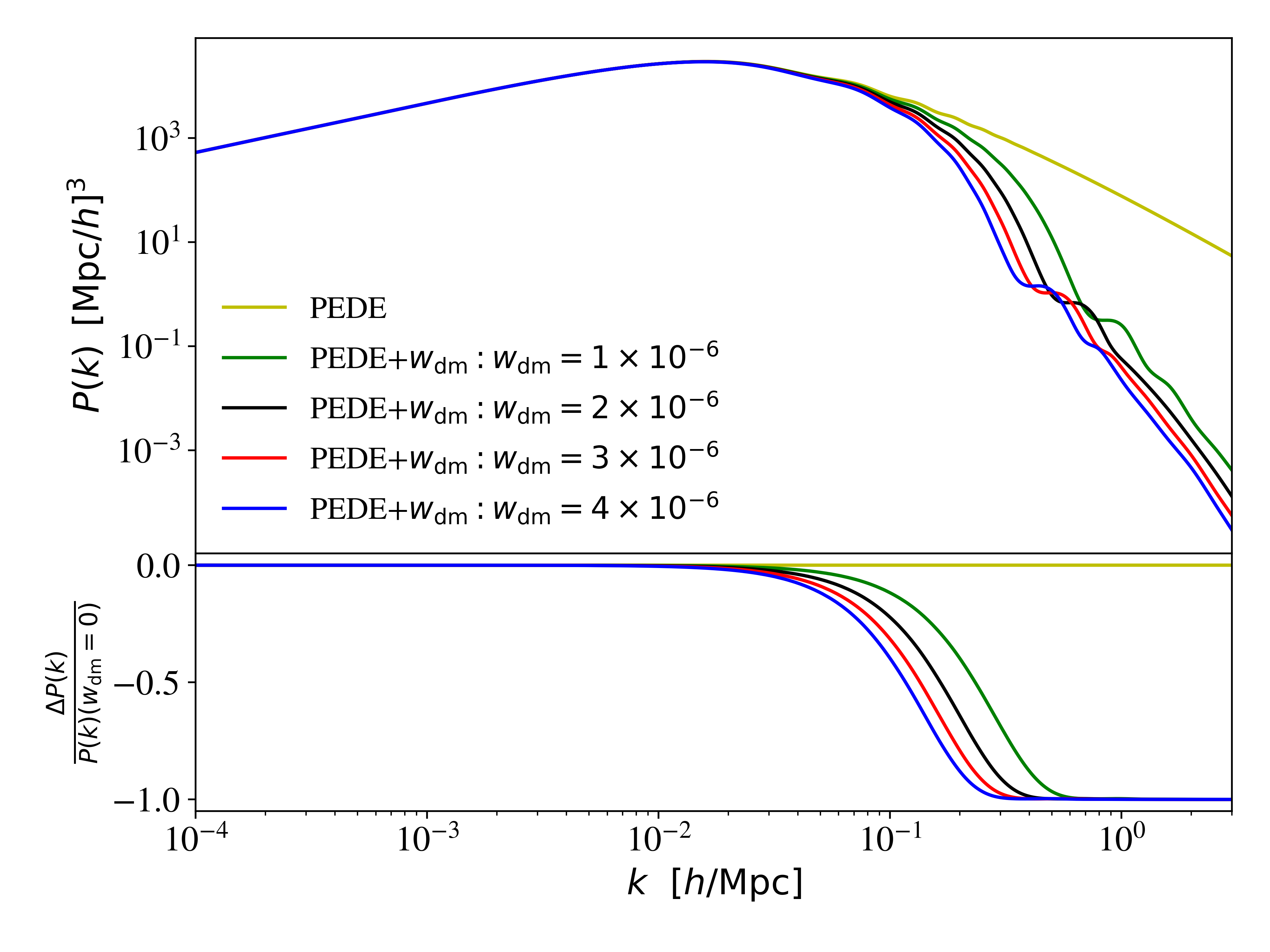}}
    \end{minipage}
    \caption{The CMB TT and matter power spectra for different values of parameter $w_{\mathrm{dm}}$, where other relevant model parameters are fixed to their bestfit values extracted from CMB+BAO+Pantheon data analysis.}
    \label{fig:CMB}
\end{figure*}

At the conclusion of this section, we provide an analysis of the effects of the PEDE+$w_{\rm dm}$ model on the CMB TT and matter power spectra across various values of $w_{\rm dm}$. Fig.~\ref{fig:CMB} illustrates the CMB TT by setting $w_{\rm dm}=1\times10^{-4}$, $2\times10^{-4}$, $3\times10^{-4}$, $4\times10^{-4}$ and matter power spectra by setting $w_{\rm dm}=1\times10^{-6}$, $2\times10^{-6}$, $3\times10^{-6}$, $4\times10^{-6}$ while keeping six other parameters fixed at their mean values derived from CMB+BAO+Pantheon data analysis.

In the context of the CMB TT power spectrum, we observe that positive values of $w_{\rm dm}$ in the PEDE+$w_{\rm dm}$ model result in a decrease in the amplitude of the acoustic peaks relative to the large-scale anisotropy. At large scales, where the integrated Sachs-Wolfe effect predominates, the primary impact of $w_{\rm dm}>0$ is an enhanced decay of the gravitational potential post-recombination up to the present time, leading to increased anisotropy for $l<100$. Since the direct effects of the DM EoS parameter on the order of $10^{-4}$ are negligible compared to those caused by a positive sound speed squared of the same order of magnitude, the effects described above are predominantly due to a positive sound speed squared rather than just a positive DM EoS parameter.

For the matter power spectrum, we can see that positive values of $w_{\rm dm}$ decrease the matter power spectrum. Similar to the case concerning the CMB TT power spectrum, such effects are mostly due to a positive sound speed squared rather than just a positive DM EoS parameter.

\section{datasets and methodology}\label{data}

To determine the free parameters of the PEDE+$w_{\mathrm{dm}}$ model, we utilize the recently described observational datasets outlined below.

\textbf{Cosmic Microwave Background (CMB)}: In this work, we employ the Planck 2018~\citep{aghanim2020planck-1,aghanim2020planck-2} data on CMB. Specifically, we utilize the CMB temperature and polarization angular power spectra plikTTTEEE+low$l$+lowE. Additionally, we incorporate the Planck 2018 CMB lensing reconstruction likelihood~\citep{aghanim2020planck-3} into our analysis.

\textbf{Baryon acoustic oscillation (BAO)}: BAO distance measurements from several astronomical surveys including 6dFGS~\citep{beutler20116df}, SDSS-MGS~\citep{ross2015clustering} and BOSS DR12~\citep{alam2017clustering} are considered.

\textbf{Supernovae Type Ia (Pantheon)}: We also incorporate the Pantheon catalog of type Ia supernovae, which consists of 1048 data points in the redshift range $z \in [0.01,2.3]$~\citep{scolnic2018complete}.

\textbf{Hubble constant (R22)}: We also present the most recent local measurement of the Hubble constant, which yields $H_0 = 73.04 \pm 1.04$ and was obtained by the SH0ES Collaboration~\citep{riess2022comprehensive}. Hereafter, we will refer to this measurement as R22.

\textbf{KiDS+VIKING-450}: We will also incorporate data from the combination of KiDS and the VISTA Kilo-Degree Infrared Galaxy Survey (VIKING), abbreviated as KV450~\citep{hildebrandt2020kids+}.

To constrain the PEDE+$w_{\mathrm{dm}}$ model, we run a Markov Chain Monte Carlo (MCMC) using the public code \textbf{MontePython-v3}~\cite{audren2013conservative,brinckmann2019montepython} and a modified version of the \textbf{CLASS} code~\cite{lesgourgues2011cosmic,blas2011cosmic}. We perform the analysis with a Metropolis-Hasting algorithm and consider chains to be converged using the Gelman-Rubin~\cite{gelman1992inference} criterion $R-1<0.03$. In Table~\ref{tab:prior} we display the priors on the free parameters of PEDE+$w_{\mathrm{dm}}$, that are, the baryon density $\omega_{b}$, the GDM density $\omega_{\mathrm{dm}}$, the ratio of the sound horizon to the angular diameter distance $\theta_s$, the amplitude and the spectral index of the primordial scalar perturbations $A_s$ and $n_s$, the optical depth $\tau_{\mathrm{reio}}$, and the GDM EoS parameter $w_{\mathrm{dm}}$.
\begin{table}[!t]
    \centering
    \renewcommand\arraystretch{1.2}
    \setlength{\tabcolsep}{3mm}{
        \begin{ruledtabular}
            \begin{tabular}{c c}
                Parameters               & Prior       \\
                \hline
                $100\omega{}_{b}$        & [0.8,2.4]   \\
                $\omega{}_{\mathrm{dm}}$ & [0.01,0.99] \\
                $100\theta{}_{s}$        & [0.5,2.0]   \\
                $ln[10^{10}A_{s}]$       & [2.7,4.0]   \\
                $n_{s}$                  & [0.9,1.1]   \\
                $\tau{}_{\mathrm{reio}}$ & [0.01,0.8]  \\
                $10^{6}w_{\mathrm{dm}}$  & [0,100]     \\
            \end{tabular}
        \end{ruledtabular}}
    \caption{Uniform priors on the free parameters of the PEDE+$w_{\mathrm{dm}}$ model.}
    \label{tab:prior}
\end{table}
Finally, we will analyze the performance of the improved PEDE+$w_{\mathrm{dm}}$ model in comparison to the $\Lambda$CDM model and the PEDE model. We utilized the cosmological code \textbf{MCEvidence}, developed by the authors of Ref.~\citep{heavens2017marginal, heavens2017no}, to calculate Bayesian evidence for all datasets and referred to Ref.~\citep{pan2018observational, yang2019constraints} for further discussion on this topic. The performance of a cosmological model (denoted as $B_{i}$) with respect to a reference cosmological model is quantified by the Bayes factor $B_{ij}$ (or its logarithm $B_{ij}$) of the model $M_i$ with respect to the reference model $M_j$. In Table~\ref{tab:Bij}, we employ the modified Jeffrey's scale which quantifies observational support for underlying cosmological models and compare our model with both $\Lambda$CDM and PEDE models respectively. In this paper, $j$ is $\Lambda$CDM, $i$ is PEDE or PEDE+$w_{\mathrm{dm}}$.

\begin{table}[!t]
    \centering
    \renewcommand\arraystretch{1.2}
    \setlength{\tabcolsep}{3mm}{
        \begin{ruledtabular}
            \begin{tabular}{c c}
                $\ln B_{ij}$            & Strength of evidence for model $M_i$ \\
                \hline
                $0 \leq \ln B_{ij} < 1$ & Weak                                 \\
                $1 \leq \ln B_{ij} < 3$ & Definite/Positive                    \\
                $3 \leq \ln B_{ij} < 5$ & Strong                               \\
                $\ln B_{ij} \geq 5$     & Very strong                          \\
            \end{tabular}
        \end{ruledtabular}}
    \caption{Revised Jeffreys' scale quantifies the strength of evidence for model $M_i$ compared to model $M_j$.}
    \label{tab:Bij}
\end{table}

\section{results and discussion}\label{results}

\begin{figure*}
    \centering
    \includegraphics[width=0.8\textwidth]{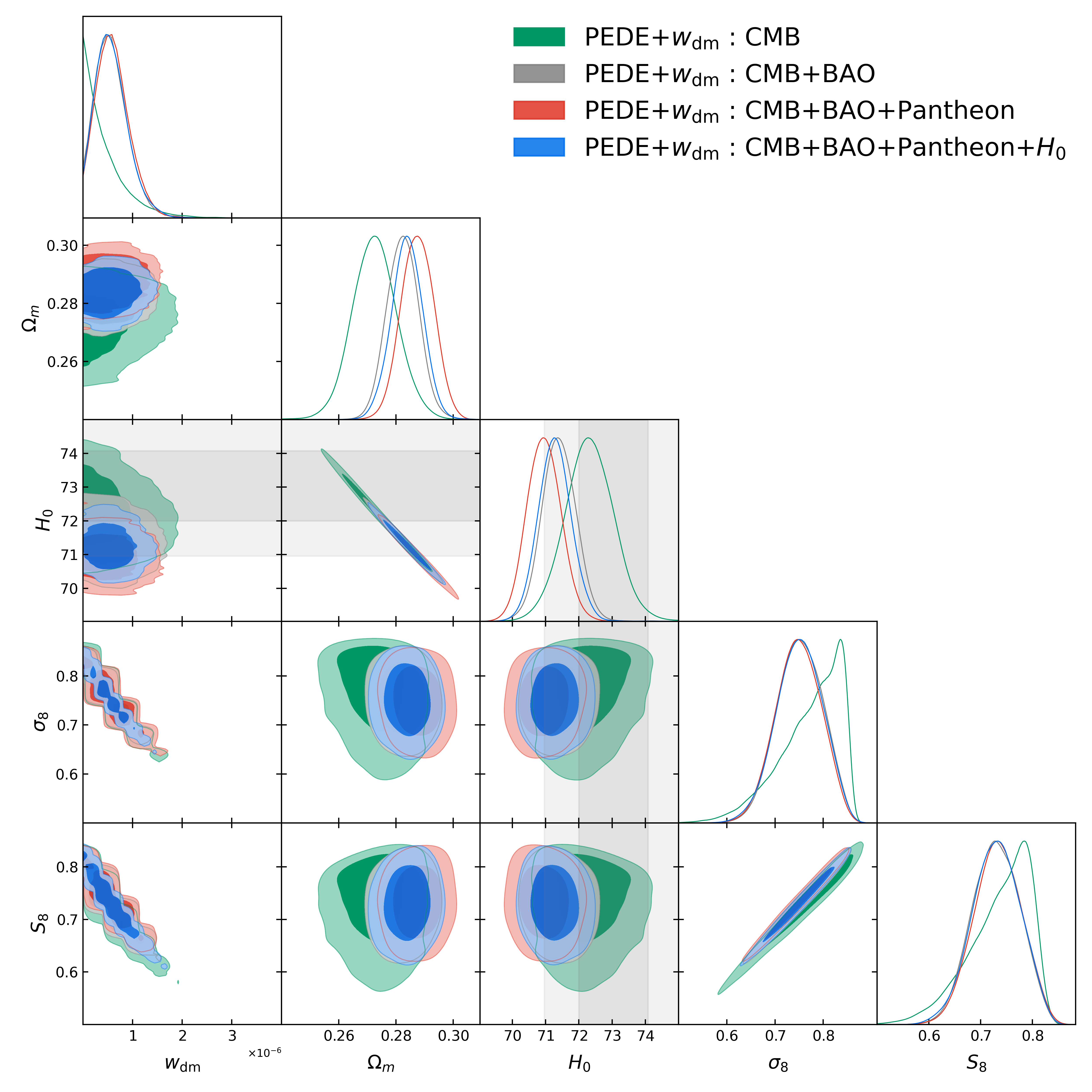}
    \caption{One-dimensional posterior distributions and two-dimensional joint contours at the 68\% and 95\% confidence levels for the most relevant parameters of the PEDE+$w_{\rm dm}$ model are shown using the CMB, CMB+BAO, CMB+BAO+Pantheon, and CMB+BAO+Pantheon+$H_0$ observational datasets. The R22 results are highlighted in grey.}
    \label{fig:pede+wdm}
\end{figure*}

In Tab.~\ref{tab:pede+wdm} and Fig.~\ref{fig:pede+wdm}, we present the constraints on the PEDE+$w_{\mathrm{dm}}$ model for CMB, CMB+BAO, CMB+BAO+Pantheon, and CMB+BAO+Pantheon+$H_0$ datasets. Additionally, we provide the best-fit values and 1$\sigma$ and 2$\sigma$ errors of the parameters of the PEDE model for the same datasets in Table.~\ref{tab:pede} for comparison.

We start by examining the fitting results of the PEDE+$w_{\rm dm}$ model using only CMB data. Subsequently, we explore the impact of incorporating additional probes by progressively adding them to the CMB analysis. Using only CMB data, we find no evidence for a non-zero DM parameter $w_{\rm dm}$ at the 68\% confidence level. Specifically, at the 68\% confidence level, $w_{\rm dm}$ ranges from $0$ to $0.539\times10^{-6}$. Nevertheless, differences in some of the model parameters between the PEDE model and the PEDE+$w_{\rm dm}$ model are presented due to the small but not negligible positive mean value of $w_{\rm dm}$. The parameters with significant changes in fitting results are $\sigma_8$ and $S_8$. More specifically, the small but non-vanish positive mean value of parameter $w_{\rm dm}$ reduces the parameter $\sigma_8$ and $S_8$ from $\sigma_8=0.856\pm0.006$ (at the 68\% confidence level) and $S_8=0.813\pm0.013$ (at the 68\% confidence level) in the PEDE model to $\sigma_8=0.771^{+0.083}_{-0.028}$ (at the 68\% confidence level) and $S_8=0.735^{+0.077}_{-0.031}$ (at the 68\% confidence level) in the PEDE+$w_{\rm dm}$ model. We point out that these changes mainly originate from the effects of a non-zero mean value of DM sound speed squared, which is on the order of $10^{-6}$, rather than directly from the effects of a DM EoS parameter of the same magnitude, because DM EoS parameter being of this magnitude has a negligible direct effect on the fitting results of other parameters. As for the other parameters, introducing a DM parameter $w_{\rm dm}$ does not cause them to undergo significant changes. Specifically, the introduction of $w_{\rm dm}$ only changes $H_0$ from $H_0=72.5\pm0.70$ (at the 68\% confidence level) in the PEDE model to $H_0=72.33\pm 0.75$ (at the 68\% confidence level) in the PEDE+$w_{\rm dm}$ model. As a result, the Hubble tension with R22 still lies within 1$\sigma$.

With the addition of BAO data to CMB, we observe some changes in the fitting results. In particular, we find an indication for a positive DM parameter $w_{\rm dm}$ at the 68\% confidence level ($w_{\rm dm}=0.63^{+0.19}_{-0.44}\times10^{-6}$ at the 68\% confidence level). Therefore, due to the positive correlation between $w_{\rm dm}$ and $\Omega_m$ along with the negative correlation between $w_{\rm dm}$ and $H_0$, $\sigma_8$, and $S_8$, we obtain a slightly higher value of $\Omega_m$ and slightly lower values of $H_0$, $\sigma_8$, and $S_8$ compared to those of the PEDE model. As a result, the $H_0$ tension with R22 is raised from 1.2$\sigma$ in the PEDE model to 1.4$\sigma$.

When Pantheon are added to CMB+BAO, we still find an indication for a positive DM parameter $w_{\rm dm}$ at the 68\% confidence level ($w_{\rm dm}=0.67^{+0.19}_{-0.45}\times10^{-6}$ at the 68\% confidence level). And because of the same reason we discussed in the previous paragraph, a slightly higher value of $\Omega_m$ and slightly lower values of $H_0$, $\sigma_8$, and $S_8$ compared to those of the PEDE model are obtained, raising the $H_0$ tension with R22 from 1.6$\sigma$ in the PEDE model to 1.8$\sigma$. We can also see that, compared to the case using CMB+BAO, the fit results for each parameter do not change significantly after adding Pantheon.

For the CMB+BAO+Pantheon+$H_0$ dataset, we continue to observe evidence for a positive DM parameter $w_{\rm dm}$ at the 68\% confidence level ($w_{\rm dm}=0.61^{+0.22}_{-0.41}\times10^{-6}$ at the 68\% confidence level). Therefore, compared to the PEDE model, there is a slight rise in $\Omega_m$ and slight declines in $H_0$, $\sigma_8$, and $S_8$, which elevates the Hubble tension with R22 from 1.3$\sigma$ in the PEDE model to 1.5$\sigma$ in our new model. It is worth mentioning that the fit results for each parameter derived from the three data sets, CMB+BAO, CMB+BAO+Pantheon, and CMB+BAO+Pantheon+$H_0$, are very similar.

\begin{figure*}
    \centering
    \includegraphics[width=0.8\textwidth]{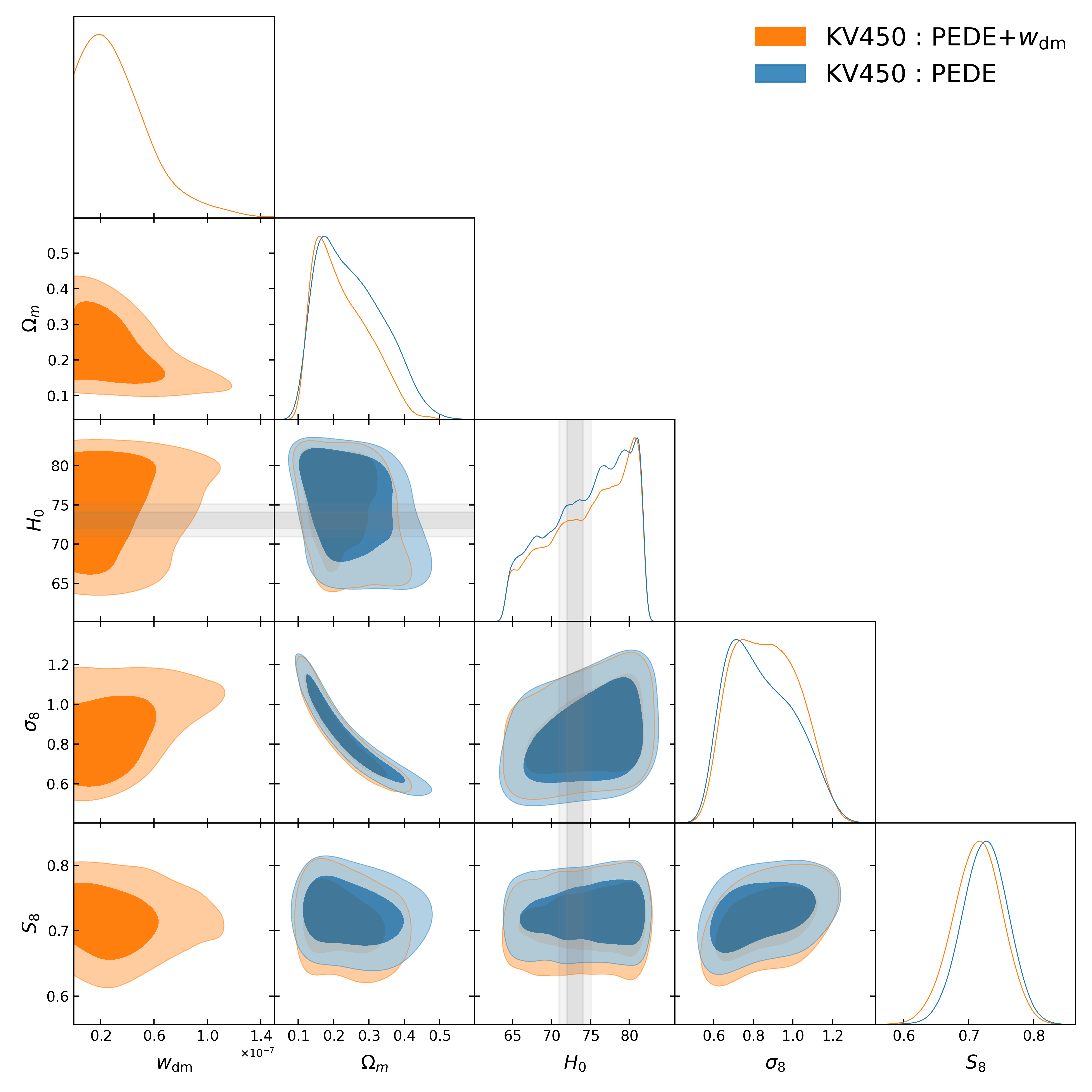}
    \caption{One-dimensional posterior distributions and two-dimensional joint contours at the 68\% and 95\% confidence levels for the most relevant parameters of the PEDE  model and the PEDE+$w_{\rm dm}$ model are shown using the KiDS+VIKING-450 dataset. The R22 results are highlighted in grey.}
    \label{fig:kv450_cf}
\end{figure*}

Considering that when using data including CMB, both $\sigma_8$ and $S_8$ are reduced compared to those of the PEDE model. Therefore, it is necessary for us to verify whether this result is consistent with the WL data, thus resolving the $\sigma_8$ and $S_8$ tension. In order to do that, we the KiDS+VIKING-450 dataset to fit the PEDE+$w_{\rm dm}$ model as well as the PEDE model, and the constraints on these two models are shown in Fig.~\ref{fig:kv450_cf} and Tab.~\ref{tab:kv450_cf}. From the Tab.~\ref{tab:kv450_cf}, we find that, for the PEDE model, the fitting results for $\sigma_8$ and $S_8$ from the KiDS+VIKING-450 dataset are $\sigma_8=0.84^{+0.13}_{-0.22}$ and $S_8=0.726\pm 0.035$ at the 68\% confidence level, respectively. Consequently, the tensions in $\sigma_8$ and $S_8$ between Planck 2018 and KiDS+VIKING-450 within the PEDE model are 0.1$\sigma$ and 2.3$\sigma$, respectively. We can see that the $S_8$ tension is on the verge of being resolved. While for the PEDE+$w_{\rm dm}$ model, the KiDS+VIKING-450 dataset yields fitting results of $\sigma_8=0.86^{+0.15}_{-0.20}$ and $S_8=0.714\pm 0.038$ at the 68\% confidence level. Accordingly, the tension between Planck 2018 and KiDS+VIKING-450 within the PEDE+$w_{\rm dm}$ model are 0.4$\sigma$ for both $\sigma_8$ and $S_8$. In summary, the new model can at least resolve the $\sigma_8$ and $S_8$ tension between Planck 2018 and the KiDS+VIKING-450 dataset.

Finally, we present the $\ln B_{ij}$ values quantifying the evidence of fit of the PEDE model and the PEDE+$w_{\rm dm}$ model with respect to the $\Lambda$CDM model under the data sets considered in this work in Tab.~\ref{tab:lnB_LCDM}. Recall the Revised Jeffreys' scale shown in Tab.~\ref{tab:Bij}, we find that although PEDE+$w_{\rm dm}$ performs better than $\Lambda$CDM when it comes to alleviating tensions in $H_0$, $\sigma_8$ and $S_8$, the Bayesian evidence shows that all the data sets considered in this work favor $\Lambda$CDM more. More specifically, except for the CMB dataset favoring $\Lambda$CDM over PEDE+$w_{\rm dm}$ with strong evidence, all other datasets favor $\Lambda$CDM over PEDE+$w_{\rm dm}$ with very strong evidence. In addition, while PEDE+$w_{\rm dm}$ outperforms PEDE in resolving tensions in $S_8$, the Bayesian evidence shows that all datasets favor PEDE over PEDE+$w_{\rm dm}$ with very strong evidence.

\begin{table*}[!t]
    \centering
    \renewcommand\arraystretch{1.5}
    \setlength{\tabcolsep}{1mm}{
        \begin{ruledtabular}
            \begin{tabular}{c c c c c}
                Parameters               & CMB                                             & CMB+BAO                                        & CMB+BAO+Pantheon                                    & CMB+BAO+Pantheon+$H_0$                               \\
                \hline
                $100\omega{}_{b}$        & ${ 2.236\pm 0.016 }^{+0.031}_{-0.030}$          & ${ 2.223\pm 0.013   }^{+0.027}_{-0.025}$       & ${ 2.216\pm 0.014 }^{+0.029}_{-0.028}$              & ${ 2.221\pm 0.013 }^{+0.026}_{-0.026} $              \\
                $\omega{}_{\mathrm{dm}}$ & ${ 0.1201\pm 0.0013}^{+0.0026}_{-0.0026}$       & ${ 0.12174\pm 0.00093 }^{+0.0018}_{-0.0018}$   & ${ 0.12256\pm 0.00094 }^{+0.0018}_{-0.0018}$        & ${ 0.12200\pm 0.00094 }^{+0.0018}_{-0.0019}$         \\
                $100\theta{}_{s}$        & ${ 1.04188\pm 0.00030 }^{+0.00058}_{-0.00059}$  & ${ 1.04173\pm 0.00030 }^{+0.00056}_{-0.00058}$ & ${ 1.04167\pm 0.00030 }^{+0.00056}_{-0.00057}$      & ${ 1.04175\pm 0.00028 }^{+0.00056}_{-0.00053}$       \\
                $\ln(10^{10}A_{s})$      & ${ 3.047\pm 0.015 }^{+0.029}_{-0.028} $         & ${ 3.041\pm 0.015     }^{+0.028}_{-0.028}$     & ${ 3.038\pm 0.017 }^{+0.031}_{-0.030}$              & ${ 3.039\pm 0.015  }^{+0.028}_{-0.028} $             \\
                $n_{s}$                  & ${ 0.9646\pm 0.0044  }^{+0.0087}_{-0.0083}$     & ${ 0.9607\pm 0.0037   }^{+0.0072}_{-0.0072}$   & ${ 0.9589\pm 0.0038 }^{+0.0073}_{-0.0074}$          & ${ 0.9602^{+0.0039}_{-0.0033} }^{+0.0068}_{-0.0076}$ \\
                $\tau{}_{reio}$          & ${ 0.0551\pm 0.0075}^{+0.015}_{-0.015}$         & ${ 0.0512\pm 0.0072}^{+0.014}_{-0.014}$        & ${ 0.0487^{+0.0082}_{-0.0068} }^{+0.016}_{-0.015} $ & ${ 0.0500\pm 0.0075     }^{+0.015}_{-0.015} $        \\
                $10^{6}w_{\mathrm{dm}}$  & $< 0.539<1.40$                                  & $ 0.63^{+0.19}_{-0.44}<1.17$                   & $0.67^{+0.19}_{-0.45} < 1.20$                       & $ 0.61^{+0.22}_{-0.41}< 1.19$                        \\
                $\Omega_m$               & ${ 0.2725\pm 0.0080  }^{+0.016}_{-0.015}$       & ${ 0.2824\pm 0.0057 }^{+0.011}_{-0.011}$       & ${ 0.2875\pm 0.0057 }^{+0.011}_{-0.011} $           & ${ 0.2839\pm 0.0056 }^{+0.011}_{-0.011} $            \\
                $H_0$                    & ${ 72.33\pm 0.75  }^{+1.5}_{-1.5}  $            & ${ 71.40\pm 0.51 }^{+1.0}_{-0.98} $            & ${ 70.95\pm 0.51   }^{+1.0}_{-0.96}$                & ${ 71.28\pm 0.50   }^{+1.0}_{-0.96} $                \\
                $\sigma_8$               & ${ 0.771^{+0.083}_{-0.028} }^{+0.094}_{-0.13}$  & ${ 0.751\pm 0.051 }^{+0.089}_{-0.093}$         & ${ 0.748\pm 0.052   }^{+0.091}_{-0.089}$            & ${ 0.750\pm 0.050}^{+0.093}_{-0.092}$                \\
                $S_8$                    & ${ 0.735^{+0.077}_{-0.031} }^{+0.096}_{-0.12} $ & ${ 0.729\pm 0.050   }^{+0.089}_{-0.086} $      & ${ 0.732\pm 0.051}^{+0.088}_{-0.090}$               & ${ 0.730\pm 0.049  }^{+0.091}_{-0.089} $             \\
            \end{tabular}
        \end{ruledtabular}}
    \caption{The mean values and 1, 2$\sigma$ of the PEDE+$w_{\mathrm{dm}}$ model for CMB, CMB+BAO, CMB+BAO+Pantheon and CMB+BAO+Pantheon+$H_0$ datasets.}
    \label{tab:pede+wdm}
\end{table*}

\begin{table*}[!t]
    \centering
    \renewcommand\arraystretch{1.5}
    \begin{ruledtabular}
        \begin{tabular}{c c c c c}
            Parameters               & CMB                                                       & CMB+BAO                                                  & CMB+BAO+Pantheon                                         & CMB+BAO+Pantheon+$H_0$                                         \\
            \hline
            $100\omega{}_{b}$        & ${2.239^{+0.015}_{-0.015}}^{+0.029}_{-0.030}$             & $ {2.227^{+0.013}_{-0.013}}^{+0.026}_{-0.026}  $         & ${2.221^{+0.013}_{-0.013}}^{+0.026}_{-0.026}$            & ${ 2.226 ^{+ 0.012 }_{- 0.012 } }^{+0.024}_{-0.025}$
            \\
            $\omega{}_{\mathrm{dm}}$ & ${0.1198^{+0.0012}_{-0.0012}}^{+0.0024}_{-0.0024}$        & ${0.1213^{+0.0009}_{-0.0009}}^{+0.0018}_{-0.0018} $      & ${0.1221^{+0.0009}_{-0.0009}}^{+0.0016}_{-0.0016}$       & ${ 0.12157 ^{+ 0.00082 }_{- 0.00082 } }^{+0.0016}_{-0.0016}$
            \\
            $100\theta{}_{s}$        & $ {1.04189^{+0.00030}_{-0.00030}}^{+0.00059}_{-0.00059} $ & ${1.04175^{+0.00030}_{-0.00030}}^{+0.00058}_{-0.00060} $ & ${1.04172^{+0.00028}_{-0.00028}}^{+0.00057}_{-0.00053}$  & ${ 1.04175 ^{+ 0.00028 }_{- 0.00028 } }^{+0.00056}_{-0.00056}$
            \\
            $\ln(10^{10}A_{s})$      & ${3.043^{+0.015}_{-0.015}}^{+0.029}_{-0.028}$             & $ {3.039^{+0.014}_{-0.014}}^{+0.027}_{-0.028} $          & ${3.036^{+0.014}_{-0.014}}^{+0.026}_{-0.027}$            & ${ 3.037 ^{+ 0.014 }_{- 0.014} }^{+0.028}_{-0.029} $
            \\
            $n_{s}$                  & ${0.9655^{+0.0042}_{-0.0042}}^{+0.0080}_{-0.0084} $       & ${0.9619^{+0.0036}_{-0.0036}}^{+0.0071}_{-0.0068} $      & ${0.9603^{+0.0035}_{-0.0035}}^{+0.0068}_{-0.0069}$       & ${ 0.9614 ^{+ 0.0035 }_{- 0.0035 } }^{+0.0069}_{-0.0072}$
            \\
            $\tau{}_{reio}$          & $ {0.054^{+0.008}_{-0.008}}^{+0.015}_{-0.014}   $         & $ {0.050^{+0.007}_{-0.007}}^{+0.014}_{-0.014} $          & ${0.048^{+0.007}_{-0.007}}^{+0.013}_{-0.014}$            & ${ 0.0494 ^{+ 0.0071 }_{- 0.0071 } }^{+0.014}_{-0.015}  $
            \\
            $\Omega_m$               & ${0.270^{+0.007}_{-0.007}}^{+0.015}_{-0.014}$             & $ {0.280^{+0.006}_{-0.006}}^{+0.011}_{-0.011}  $         & ${0.285^{+0.005}_{-0.005}}^{+0.010}_{-0.010}$            & ${ 0.2812 ^{+ 0.0049 }_{- 0.0049 } }^{+0.0096}_{-0.0092}$
            \\
            $H_0$                    & ${72.5^{+0.70}_{-0.70}}^{+1.4}_{-1.4} $                   & ${71.7^{+0.51}_{-0.51}}^{+1.0}_{-1.0}   $                & ${71.21^{+0.47}_{-0.47}}^{+0.91}_{-0.91} $               & ${ 71.52 ^{+ 0.44 }_{- 0.44 } }^{+0.84}_{-0.85}$
            \\
            $\sigma_8$               & $  {0.856^{+0.006}_{-0.006}}^{+0.012}_{-0.012}      $     & $  {0.858^{+0.006}_{-0.006}}^{+0.012}_{-0.012}  $        & ${0.859^{+0.006}_{-0.006}}^{+0.012}_{-0.012}$            & ${ 0.8577 ^{+ 0.0063 }_{- 0.0063 } }^{+0.012}_{-0.012}$
            \\
            $S_8$                    & ${ 0.813 ^{+ 0.013 }_{- 0.013 } }^{+ 0.026 }_{- 0.025 }$  & ${ 0.828 ^{+ 0.011 }_{- 0.011 } }^{+ 0.020 }_{- 0.021 }$ & ${ 0.836 ^{+ 0.010 }_{- 0.010 } }^{+ 0.020 }_{- 0.019 }$ & ${ 0.830 ^{+ 0.010 }_{- 0.010 } }^{+0.019}_{-0.020} $
            \\
        \end{tabular}
    \end{ruledtabular}
    \caption{The mean values and 1, 2$\sigma$ of the PEDE model for CMB, CMB+BAO, CMB+BAO+Pantheon and CMB+BAO+Pantheon+$H_0$ datasets.}
    \label{tab:pede}
\end{table*}

\begin{table*}[!t]
    \centering
    \renewcommand\arraystretch{1.5}
    \setlength{\tabcolsep}{3mm}{
        \begin{tabular}{c c c }
            \hline
            \hline
            Parameters               & PEDE                                                & PEDE+$w_{\mathrm{dm}}$                          \\
            \hline
            $100\omega{}_{b}$        & ${ 2.24 ^{+ 0.22 }_{- 0.33 } }^{+ 0.36 }_{- 0.35 }$                                                   \\
            $\omega{}_{\mathrm{dm}}$ & ${0.115^{+0.034}_{-0.063} }^{+0.088}_{-0.073}$      & ${ 0.102^{+0.025}_{-0.054} }^{+0.077}_{-0.061}$ \\
            $ln[10^{10}A_{s}]$       & ${ 3.10^{+0.61}_{-1.3}  }^{+1.7}_{-1.4} $           & ${ 3.25^{+0.79}_{-1.2} }^{+1.6}_{-1.5}$         \\
            $n_{s}$                  & ${ 1.04^{+0.16}_{-0.12}    }^{+0.24}_{-0.26} $      & ${ 1.14^{+0.15}_{-0.052}  }^{+0.17}_{-0.25} $   \\
            $10^{8}w_{\mathrm{dm}}$  & -                                                   & $< 4.27< 8.69$                                  \\
            $\Omega_m$               & ${ 0.250^{+0.064}_{-0.12}  }^{+0.17}_{-0.14}$       & ${ 0.225^{+0.048}_{-0.10}  }^{+0.15}_{-0.12} $  \\
            $H_0$                    & ${ 74.8^{+7.0}_{-2.8} }^{+7.3}_{-9.1}$              & ${ 74.9^{+6.9}_{-2.8}   }^{+7.2}_{-9.7}$        \\
            $\sigma_8$               & ${ 0.84^{+0.13}_{-0.22} }^{+0.32}_{-0.27}  $        & ${ 0.86^{+0.15}_{-0.20}  }^{+0.30}_{-0.28} $    \\
            $S_8$                    & ${ 0.726\pm 0.035  }^{+0.067}_{-0.070} $            & ${ 0.714\pm 0.038   }^{+0.072}_{-0.078}$        \\
            \hline
            \hline
        \end{tabular}}
    \caption{The mean values and 1, 2$\sigma$ of the PEDE+$w_{\mathrm{dm}}$ model for the KiDS+VIKING-450 dataset.}
    \label{tab:kv450_cf}
\end{table*}

\begin{table}[!t]
    \centering
    \begin{ruledtabular}
        \begin{tabular}{l c r}
            Model                  & Datasets               & $\ln B_{ij}$ \\
            \hline
            PEDE                   & CMB                    & 0.93         \\
            PEDE                   & CMB+BAO                & -2.32        \\
            PEDE                   & CMB+BAO+Pantheon       & -8.02        \\
            PEDE                   & CMB+BAO+Pantheon+$H_0$ & 2.00         \\
            PEDE                   & KiDS+VIKING-450        & -0.002       \\
            \\
            PEDE+$w_{\mathrm{dm}}$ & CMB                    & -4.37        \\
            PEDE+$w_{\mathrm{dm}}$ & CMB+BAO                & -10.80       \\
            PEDE+$w_{\mathrm{dm}}$ & CMB+BAO+Pantheon       & -17.00       \\
            PEDE+$w_{\mathrm{dm}}$ & CMB+BAO+Pantheon+$H_0$ & -8.06        \\
            PEDE+$w_{\mathrm{dm}}$ & KiDS+VIKING-450        & -6.94        \\
        \end{tabular}
    \end{ruledtabular}
    \caption{Summary of the $\ln B_{ij}$ values, which quantify the evidence of fit for the PEDE model and the PEDE+$w_{\rm dm}$ model relative to the $\Lambda$CDM model, using the CMB, CMB+BAO, CMB+BAO+Pantheon, CMB+BAO+Pantheon+$H_0$, and KiDS+VIKING-450 datasets. It should be noted that a negative $\ln B_{ij}$ value indicates that the PEDE model or the PEDE+$w_{\rm dm}$ model is less supported compared to the base $\Lambda$CDM model.}
    \label{tab:lnB_LCDM}
\end{table}

\section{concluding remarks}
\label{conclusion}
Although CDM is supported by a wide range of cosmological data when it is accompanied by a cosmological constant, it encounters many issues inherent to the CDM paradigm at small scales, such as the missing satellite, too-big-to-fail, and core-cusp problem. These issues have motivated many new DM candidates beyond CDM, including warm DM, fuzzy DM, interacting DM and decaying DM, which are all proposed to suppress the formation of low-mass structures. The current limitations of CDM leave the question of "What is DM?" relatively open. In light of this, we have proposed a new cosmological model by considering DM as a barotropic fluid characterized by a constant EoS parameter and DE as PEDE, considering that there have already been many researches on the extended properties of DM in the context of a cosmological constant and there are some interesting results emerged as we investigated the extended properties of DM in the context of PEDE.~\citep{yao2024observational}.

Given the basic equations of the PEDE+$w_{\rm dm}$ model both at the background and perturbation levels, we present some analysis concerning the impacts of the PEDE+$w_{\rm dm}$ model on the CMB TT and matter power spectra for different values of $w_{\rm dm}$, and find that positive values of $w_{\rm dm}$ result in a reduction in the amplitude of the acoustic peaks compared to the large-scale anisotropy occurred, and an increase in the power at large scale where $l<100$ due to integrated Sachs-Wolfe effect. In addition to that, positive values of $w_{\rm dm}$ decrease the matter power spectrum. We pointed out here that the effects on both CMB TT and matter power spectra are mostly due to a positive sound speed squared rather than just a positive DM EoS parameter. We have also fitted the model using data sets consisting of CMB, CMB+BAO, CMB+BAO+Pantheon, CMB+BAO+Pantheon+$H_0$, and KiDS+VIKING-450. We find that the value of parameter $w_{\rm dm}$ is positive at 68\% confidence level for CMB+BAO, CMB+BAO+Pantheon and CMB+BAO+Pantheon+$H_0$ datasets, which leads to a slight rise in $\Omega_m$ and slight declines in $H_0$, $\sigma_8$, and $S_8$. Consequently, the $H_0$ tension with R22 is raised from 1.2$\sigma$ (CMB+BAO), 1.6$\sigma$ (CMB+BAO+Pantheon), and 1.3$\sigma$ (CMB+BAO+Pantheon+$H_0$) in the PEDE model to 1.4$\sigma$ (CMB+BAO), 1.8$\sigma$ (CMB+BAO+Pantheon), and 1.5$\sigma$ (CMB+BAO+Pantheon+$H_0$) in the PEDE+$w_{\rm dm}$ model. Of course, these levels of tensions still can be attributed to the statistical fluctuations. We also find that the $S_8$ tension between the Planck 2018 data and the KiDS+VIKING-450 data are reduced from 2.3$\sigma$ in the PEDE model to 0.4$\sigma$ in the new model. Finally, it is also noteworthy that, although PEDE+$w_{\rm dm}$ outperforms PEDE in addressing $S_8$ tension between CMB and KiDS+VIKING-450, Bayesian evidence suggests that PEDE favors our new model with very strong evidence by all the datasets considered in this study. Therefore, we conclude in the end that the PEDE+$w_{\rm dm}$ model is not a good alternative to the PEDE model.
\begin{acknowledgments}
    This work is supported by the National key R\&D Program of China (Grant No.2020YFC2201600), Guangdong Basic and Applied Basic Research Foundation (Grant No.2024A1515012573), National Natural Science Foundation of China (Grant No.12073088), and National SKA Program of China (Grant No. 2020SKA0110402).
\end{acknowledgments}

\bibliography{bibtex}

\end{document}